\documentclass[conference]{IEEEtran}
\IEEEoverridecommandlockouts
\usepackage{blindtext, graphicx}

\usepackage[usenames,dvipsnames]{pstricks}
\usepackage{epsfig}
\usepackage{moredefs}
\usepackage{etex}
\usepackage{amsmath,amssymb,amsthm,mathtools, placeins,multirow,booktabs,fixltx2e,rotating,graphicx,color}
\usepackage[linesnumbered,algoruled,vlined]{algorithm2e}

\usepackage[justification=centering]{caption}

\usepackage{filecontents}
\usepackage[noadjust]{cite}
\usepackage{multirow}
\usepackage{enumerate}
\usepackage{ucs,enumerate}
\usepackage{multirow}
\usepackage{booktabs}
\usepackage{color,soul}
\usepackage{caption}
\usepackage{subcaption}
\usepackage{url}

\def\BibTeX{{\rm B\kern-.05em{\sc i\kern-.025em b}\kern-.08em
    T\kern-.1667em\lower.7ex\hbox{E}\kern-.125emX}}
\begin{document}

\title{On the design of a Fog computing-based, driving behaviour monitoring framework \\
\thanks{Part of this work has been supported by the H2020-ICT-24-2016 project GamECAR (Grant No. 732068) and the H2020-SC1-DTH-2018-1 project SmartWork (Grant No. 826343).}
}

\author{\IEEEauthorblockN{Dimitrios Amaxilatis, Christos Tselios$^1$, Orestis Akrivopoulos and Ioannis Chatzigiannakis$^2$} \\
\IEEEauthorblockA{\small SparkWorks ITC Ltd, United Kingdom,
\{d.amaxilatis, akribopo\}@sparkworks.net \\
\small $^1$ Dept. of Electrical and Computer Engineering, University of Patras, Greece,  tselios@ece.upatras.gr \\
\small $^2$ Sapienza University of Rome, Italy, ichatz@diag.uniroma1.it 
}
}

\maketitle

\begin{abstract}

Recent technological improvements in vehicle manufacturing may greatly improve safety however, the individuals' driving behaviour still remains a factor of paramount importance with aggressiveness, lack of focus and carelessness being the main cause of the majority of traffic incidents. The imminent deployment of 5G networking infrastructure, paired with the advent of Fog computing and the establishment of the Internet of Things (IoT) as a reliable and cost-effective service delivery framework may provide the means for the deployment of an accurate driving monitoring solution which could be utilized to further understand the underlying reasons of peculiar road behaviour, as well as its correlation to the driver's physiological state, the vehicle condition and certain environmental parameters. This paper presents some of the fundamental attributes of Fog computing along with the functional requirements of a driving behaviour monitoring framework, followed by its high level architecture blueprint and the description of the prototype implementation process. 
\end{abstract}


\section{Introduction}
The imminent deployment of 5G networking infrastructure and the tremendous boost in coverage and performance it proclaims \cite{bianchi2016superfluidity}, along with the establishment of the Internet of Things (IoT) as a reliable and cost-effective service delivery framework, will unlock new and exciting verticals with a significant impact in our daily lives. Automotive industry is one of the markets that will be greatly benefited by the advent of 5G and the new levels of quality of experience (QoE) it introduces \cite{tselios2016camadqoe}. Road safety and traffic efficiency services will be upgraded through seamlessly interconnected devices and advanced V2X communication schemes \cite{vlachos5Gcamad2017}, while latency decrease will most likely allow semi-autonomous driving to become a commodity available to everyone. The specific vertical contributes to huge societal and economical impact, since it may render severe traffic accidents, increased energy consumption and long commute times obsolete. 

Even though technological innovation in vehicle hardware and software greatly improves safety, a person's driving behaviour remains a factor of paramount importance. Aggressiveness, lack of focus and carelessness cause many traffic incidents, while novice drivers often get involved in hazardous situations on the road. Despite its significance, there is no simple metric for quantifying aggressiveness or characterizing driving styles \cite{nousias2018}. Current attempts have either tried to pre-define characteristics of specific driving behaviours or to cluster similar driving patterns but due to the large amount of generated data online knowledge discovery techniques are necessary for extracting valuable information \cite{nousias2019chapter}. Moreover, the circumstances of driver aggressiveness must be examined under the prism of his physiological status. The integration of both in-vehicle data as well as the physiological data of the driver introduce challenges in determining the overall driving style. It is therefore essential for any contemporary sensing system to focus on determining the most influential factors, through a set of appropriate sensors that allow the driver to retain road perception \cite{nousias2018camad}, while also taking into account the given driving location.

The deployment of an end-to-end system for obtaining driver and vehicle data, execute specific analysis based on pre-defined algorithms to extract information that can be utilized to monitor and improve one's driving behaviour has always been a topic of active research. Alas until recently sensing infrastructure had not been adequately evolved to align with the real-world requirements of such a platform, while the necessary communication and networking architectural structural elements that could potentially allow the implementation of a holistic framework were put together with the advent of Fog computing \cite{chiang-fog-iot-2016}.  

The rest of the paper is organized as follows: Section II presents some of the essential attributes of Fog computing along with its benefits towards deploying an end-to-end, sensor-based platform. Section III focuses on the realistic requirements and the architecture of a sophisticated driving behaviour monitoring framework, while Section IV describes the actual prototype implementation challenges and evaluation process. Finally, Section V draws conclusions and summarizes the paper.

\section{Fundamental Attributes of Fog Computing}
\label{sec:fog-comp-attributes}


Sensing nodes tend to be physically located close to the phenomenon they monitor while their most common deployment method is over wide-area network topologies. This introduces a severe communication overhead with the back-end data centers and inevitably dictates the introduction of an intermediate intelligence layer between these entities specifically designed to eliminate large round-trip delay and possible network congestion. Such a layer allows the deployment of latency-sensitive applications and further augment the overall performance of the network. The preeminent design guideline of the aforementioned layer was to move computation resources closer to the end-user domain, in an attempt to facilitate data processing and manipulation on the spot thus eliminating the need for transmitting bulk datasets across the entire topology. This new concept is often referred to as as \textit{Edge computing} and constitutes an improved version of the existing edge network.


Multi-access edge computing (MEC) \cite{mec-samdanis} and Fog computing are considered the prevailing deployment blueprints amongst the several edge-centric computing paradigms proposed by industry and academia. MEC architecture dictates a combined operation of dedicated servers placed on the network edge, paired with cellular base stations and specific communication interfaces toward the back-end cloud infrastructure. This model appears to be mostly suitable for large scale telco offerings since it is primarily focused on network efficiency through agile and highly adaptive initiation of cellular-based services \cite{bolivar2018}. To the contrary, Fog computing appears to be more focused on real-world IoT deployment requirements by engaging both edge and core networking components as computational infrastructure, thus allowing a huge number of sensors/devices to be simultaneously monitored \cite{fog-compsac}. As a consequence, multi-tier application deployment becomes easier, obtained datasets are stored and processed closer to the original source \cite{fog-etfa, akrivopoulos2018icc}, leading to minimized service delivery latency which is essential in real-time and near real-time use cases.


Designed primarily as a distributed paradigm strategically placed between the cloud and the sensing nodes, Fog incorporates dedicated communication interfaces with the network backend. This enhances the overall topology robustness since ingress packets undergo a secondary inspection progress capable of identifying problematic or malicious content, way before reaching the cloud entry point. Fog resolves additional IoT-related constraints such as (i) the extensive bandwidth requirements due to the higher number of interconnected devices and (ii) the elevated propagation error due to the increased volume of transmitted data. However, the major contribution of Fog is without a doubt latency elimination which renders the deployment of delay-critical services possible.

\section{Functional Requirements and System Architecture}
\label{sec:req-architecture}

A generic categorization of entities necessary for implementing a driving behaviour monitoring framework which will unobtrusively record physiological, behavioural, environmental and vehicle parameters becomes clear by reviewing the core design principle of Fog computing which dictates the introduction of an intermediate layer between sensing nodes and backend infrastructure.

\begin{figure}[ht]
\centering
\includegraphics[width=0.47\textwidth]{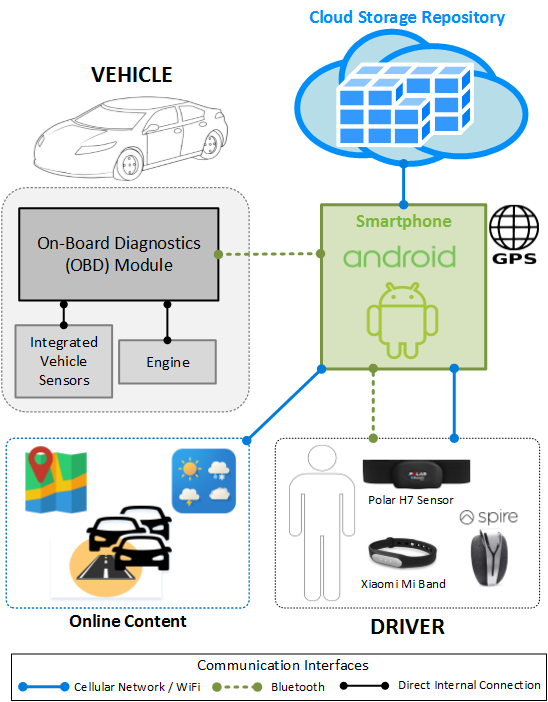}
\caption{High-level Architecture of the Driving Behaviour Monitoring Framework}
\label{architecture}
\end{figure}

As shown in Figure \ref{architecture} there are three categories of sensing devices and information retrieval services each contributing with specific bits of information necessary to fill all gaps and assemble the overall context of each route. 

\begin{figure*}[ht]
\includegraphics[width=14cm, height=5cm]{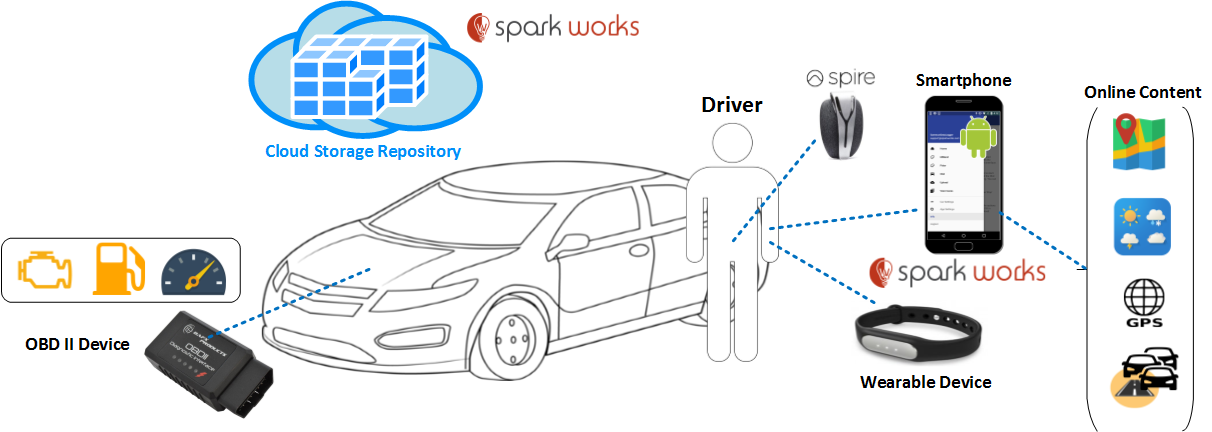}
\centering {\caption{Deployment of the driving behaviour monitoring framework prototype \label{deployment}}}
\centering
\end{figure*}

\begin{enumerate}
    \item \textbf{Vehicle sensors} are integrated to modern cars by all manufacturers. Such sensors monitor gear, tire pressure, temperature and oil, all interlinked through an internal controller area network (CAN) designed to allow seamless and robust communication. Data is sent to the vehicle's electronic control unit (ECU) and can be retrieved using an on-board diagnostics (OBD) controller, via the Bluetooth protocol. OBD exposes coded information which may slightly vary on each vehicle, however all available parameters are addressed by unique identification numbers. Through the OBD it is possible to get a detailed log of the vehicle's behaviour on any given time, which if properly time-stamped can be easily correlated with supplementary logs from other sources. 
    \item \textbf{Wearable Devices} are attached to the driver's body and record certain aspects of its physiological condition through an array of sensors in a non intrusive manner. Some contemporary wearables are also able to provide instant feedback to the driver and analyze traces collected from the sensors on the spot. The original traces (raw data) after being encrypted, may be stored locally on the internal memory of the wearable device and constitute a type of short-term inventory that will be later be further processed.
    \item \textbf{Online data repositories} containing all types of real-time data are nowadays virtually omnipresent and can be accessed for free. This allows third-party frameworks to retrieve information about the weather, traffic congestion in a specific area as well as detours and road blocks that may have impact on the duration of an individual's daily commute. Such datasets when associated with vehicle and physiological metrics may reveal driving behaviour patterns that otherwise lack of proper explanation.
\end{enumerate}

In order to align with the Fog computing design guidelines it is possible to use an Android\footnote{https://www.android.com/} smartphone as the main coordinating node of the proposed framework, resembling to the smart gateway often mentioned in similar deployments \cite{fog-compsac}. This is achieved through a custom-made application capable of connecting to all available sensors as well as third-party online repositories, collect and locally store data before transmitting them to any permanent storage repository.

Selecting a smartphone as the coordination node of any Fog-based data retrieval and processing platform provides significant flexibility due to the large number of communication protocols and corresponding interfaces any contemporary device supports, as well as the inherent caching and processing capabilities it incorporates. In addition, the accumulated datasets may undergo data pre-processing, customized for extracting the most essential and meaningful information or fill possible voids which may lead to inaccurate patterns and results as described in \cite{nousias2018}. Data pre-processing techniques or algorithms for tackling nonuniformities are relatively easy to be implemented and then integrated in the pipeline of data handling of every contemporary smartphone operating system, thus providing additional benefits after eliminating the overhead of modifying complex or proprietary software running in routers or switches.

Through this coordination node, all accumulated datasets after being processed or partially analyzed will be uploaded to an affiliated cloud storage repository, which optionally may be upgraded by dedicated processing resources rendering it capable of processing and analyzing large data sets through custom algorithms in the most efficient manner. It is also important this repository to incorporate cutting-edge security and data leakage prevention mechanisms, given the fact that some datasets may contain sensitive medical information which is often subdue to specific legislation.


\section{Prototype Implementation and Evaluation}
\label{sec:prototype-implementation}

\begin{figure*}[ht]
\includegraphics[width=17cm, height=4.5cm]{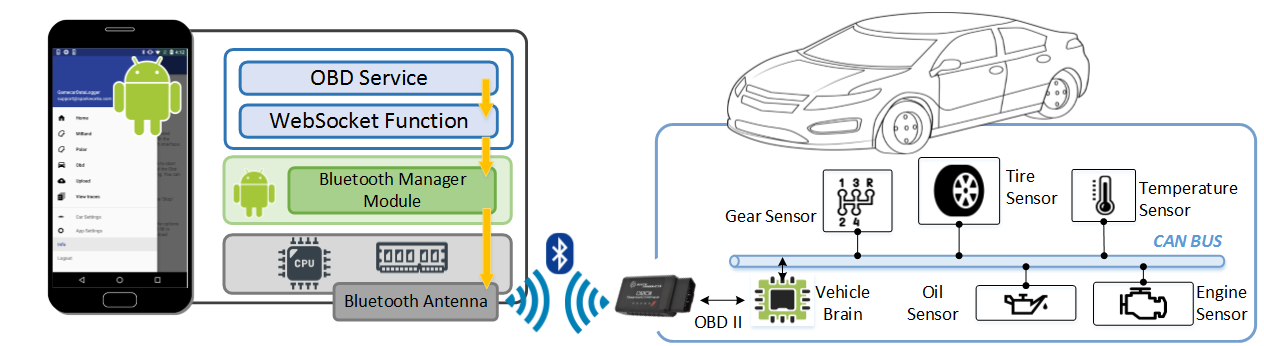}
\centering {\caption{Integrating Data Logger application and the OBD module for obtaining vehicle information \label{obd-can}}}
\centering
\end{figure*}

\par For properly evaluate the previous categorization, extract certain results on how Fog-enabled ecosystems can be seamlessly deployed and validate the proposed architecture of Section \ref{sec:req-architecture} we have implemented a functional prototype of a driving behaviour monitoring framework and conducted several experimentation routes. The fundamental components of the prototype include (i) Wearables (ii) Vehicle-sensing equipment (iii) online data repository retrieval mechanisms (iv) the Data Logger\footnote{https://play.google.com/store/apps/details?id=app.gamecar.\\sparkworks.net.gamecardatalogger}, a customized Android application for converting the driver's smartphone to a central communication and data aggregation hub and (v) the Spark Works Cloud Storage Repository which handles data storage and allows second-phase access.      

More specific, vehicle-oriented data collected by the embedded sensors of the car were obtained using an On-Board Diagnostics (OBD) module, supported by every major manufacturer following specific EU regulatory guidelines. Without the loss of generality, only data regarding vehicle speed, engine rounds-per-minute (RPM) and throttle position were collected for creating matrices stored in a per-trip fashion. The OBD module was connected over Bluetooth protocol to the driver's Android smartphone, on which Data Logger application was running. The driver's wearables were also providing data to the Data Logger after being paired and identified by the later. Data Logger acted as a data aggregator that accumulated sensor values, added a timestamp and created a .CSV file. This file was also populated by additional content retrieved from the affiliated online sources, as well as the smartphone's integrated GPS which indicated the exact positioning of the driver/vehicle. At the end of each route, the application after notifying the driver, encrypts and transmits the .CSV file through WiFi or 4G/LTE networks. Additional implementation information per group is listed below.



\subsection{On-Board Diagnostics module}
The prototype framework uses the OBD module to retrieve data from the integrated vehicle sensors. Figure \ref{obd-can} presents the implemented software stack that works in tandem with the underline hardware resources to fetch the available vehicle information. More specific, the Data Logger application contains a dedicated function, called \textit{OBD\_Service} which triggers the WebSocket function allowing to establish a direct line of communication between the application and the OBD over the integrated Bluetooth antenna. More specific, the WebSocket function, initiates the Bluetooth Manager Module of the Android Operating system which then activates the necessary hardware ports for having the Bluetooth antenna establish the necessary channel with the OBD.

\begin{figure}[ht]
\centering
\includegraphics[width=0.47\textwidth]{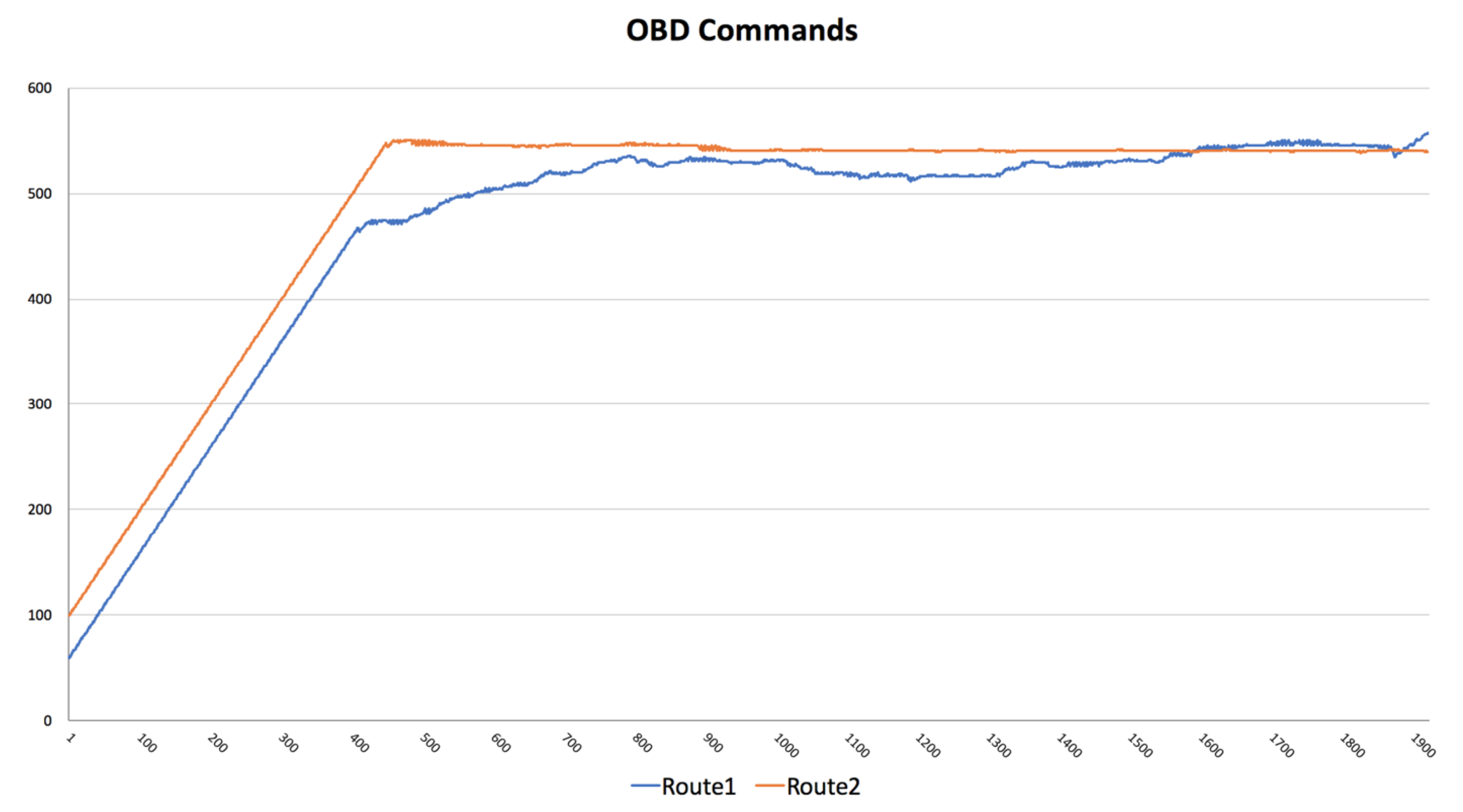}
\caption{OBD recordings}
\label{fig:obd-rec}
\end{figure}

On the other side, the OBD device, after being attached to the Serial Port existing in every vehicle, "translates" egress messages coming from the vehicles' CAN bus. This communication is amphidromous, with the OBD also pushing requests (in the form of OBD commands) for sensor information towards the vehicle's brain which are accommodated in due time (in the form of OBD command responses). The accommodation time interval varies and is dependent on the vehicle manufacturer as well as the vehicle model. After properly analyzing trip logs from several different manufacturers, we estimate that the average reply delay per OBD command is approximately 110ms. The reply timeframe per OBD command according the protocol design documentation, spans between 50ms and 200ms. These metrics indicate a maximum of 1200 OBD command replies per minute and a minimum of 300 OBD commands. The average rate yielded in our experiments was a result of a reply rate of 540 OBD commands per minute. 
For properly evaluating the described implementation, we conducted several we different driving sessions with numerous vehicles and different drivers all returning similar metrics. Figure \ref{fig:obd-rec} presents data from two different 5-minute drives occurred over two consecutive days. The X-Axis represents the number of updates received by the system in each 5-minute driving session, while Y-Axis shows the number of OBD commands recorded in the specific update. As shown, the application demonstrates a sharp increase on the number of recorded OBD commands during the first 400-500 updates, followed by a lower increase during the next 150 updates. After this, the number of OBD commands recorded on each update stabilizes at approximately 540 OBD commands per update.

\subsection{Wearable Devices}
As already stated, wearables are non-intrusive devices which record certain aspects of the physiological conditions of the driver through an array of sensors. In the specific prototype, three different wearables were utilized to obtain data regarding the driver's heartbeat and respiration rate, both factors of significant value to determine stress and anxiety on any given moment. All traces and the alerts produced along with their timestamps are initially stored in the internal memory of the device and in certain time intervals are being forwarded to the Data Logger application.
\subsubsection{Xiaomi MiBand M1S}
The Xiaomi MiBand is a wearable activity tracker consists of the core tracker which is around 9 mm thick, and 36 mm in length, inserted into a hypoallergenic TPSiV wristband, having anti-UV and anti-microbal properties. The tracker was used to access heart rate information limited to beats per minute and can offer up to an average of 1 measurement per 10 seconds due to its operation limitations and its on-demand measuring system, which is implemented based on integrated Bluetooth communication. 
\subsubsection{Polar H7 Respirator}
Polar H7 Heart Rate sensor is a device mostly used to access heart rate information and includes beats per minute as well as R-R intervals for the heartbeats. The measurement rate is around 1 measurement per 2 second as its operation is subscription based (using Bluetooth Low Energy (BLE) 4.0 subscriptions).
\subsubsection{Spire Respirator}
Spire Respirator is a wearable stress and activity tracker worn on the waistband or bra strap designed to analyze breath rates to determine levels of tension, calm, or focus. Data Logger application provides support for the Spire Respirator sensor and also integrates a flow for accessing the corresponding web platform through the provided API.

\subsection{Data Logger Application}
Data Logger is installed on the driver's mobile Android device and is paired to the available wearables for trace acquisition. During the application instatiation, the driver must pair the mobile device with the wearables following the standard BLE bonding process. As soon as the pairing process is complete, the mobile application locks the wearable device preventing it from being paired with another mobile device. User information stored within the wearables are protected from being accessed without permission even from the driver's his own device by the available mechanisms of the Android operating system. 

The application communicates with the wearable devices over a well defined API via the secure Bluetooth wireless connection, can retrieve the traces and alerts either in small packages or in batch mode and the data received are stored within the mobile device's internal storage space. The mobile application can erase some or all of the data stored (a) on the wearable device and (b) on the internal store of the mobile device.

The application is capable of analyzing the data retrieved from the medical device by utilizing a series of algorithms available for Android OS or through tailor-made ones. As data is received from the wearables, the Alert Handling component is activated to process and analyze the data and provide alerts. Data collected from the device and produced by the Alert Handling component is stored in the Data Handling component and complement those produced by the algorithms executed by the wearable device. Apart from the data transfer and management, the mobile application supports configuration/personalization tasks for the wearable device related to the memory (e.g., clean), alert generation and algorithm parametrization, battery configuration, sensors and synchronization functionality. Moreover, Data Logger also incorporates functions for obtaining traces from sensors or additional sources provided by the smartphone, such as the integrated GPS or the accelerometer. Such traces are combined with the rest to provide a holistic route overview containing the full spectrum of available information.  


\begin{figure}[ht]
\centering
\includegraphics[width=4.5cm, height=7.5cm]{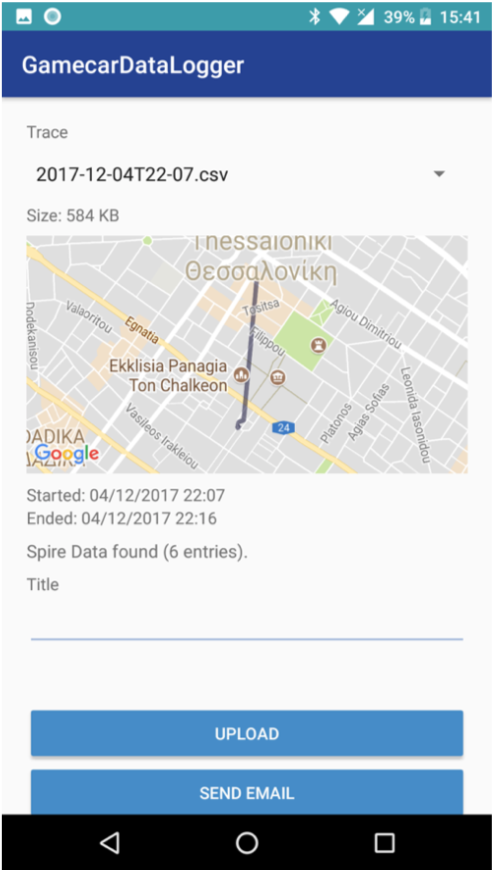}
\caption{Driving session trace upload using the Data Logger app}
\label{fig:applic-trace-upload}
\end{figure}

\subsection{Cloud Storage Repository}
All trace files holding the combined vehicle-oriented, wearable sensor data are stored in the SparkWorks Cloud Storage Repository. This repository is designed and impemented by SparkWorks to store content in a specially designed filesystem with a hierarchical structure, utilizing advanced hashing techniques for seamless data retrieval.
To provide a contemporary, efficient and scalable way to upload data trace files, SparkWorks Cloud Storage Repository provides a public REST API protected via another module of the overall SparkWorks Cloud Platform, the SparkWorks Authentication and Authorization Infrastructure which serves the trace file upload requests reliable and efficiently via multipart file upload. Upon the reception of a trace file the Cloud Storage Repository API persists the trace file metadata in a relational database. After successfully storing the trace file in the Cloud Storage Repository a unique public reference is returned to the client which uploaded the trace in the first place. At this point, the trace is already available in the Cloud Storage Repository and every authenticated client with the right permissions (as defined by the SparkWorks Authentication and Authorization Infrastructure scheme) can retrieve the trace file via a Spark Works Core REST API endpoint. The Core API is responsible to recover the stored trace file from the SparkWorks Cloud Storage Repository and make it available to the client along with the trace metadata.

\subsection{Third-party Online Applications and Cloud Services}
\subsubsection{Online Traffic Flow}
Data Logger application integrates the Online Traffic Flow, an online suite of web services for developers to create web and mobile applications around real-time traffic. The aforementioned services can be used via RESTful APIs, while the Online Traffic Flow API’s are based on real-time traffic data\footnote{https://www.tomtommaps.com/livetraffic/} with accurate and fresh information about traffic jams, incidents and flow. This service is based on flow segment data, which provides information about the speeds and travel times of the road fragment closest to any given coordinates. It is designed to work alongside the integrated Maps API to support clickable flow data visualizations. With this API, the client side can connect any place in the map with flow data on the closest road and present it to the user. 
\subsubsection{OpenWeatherMap}
OpenWeatherMap\footnote{ https://www.openweathermap.org/} is an online service that provides weather data, including current weather data, forecasts, and historical data to the developers of web services and mobile applications. For data sources, it utilizes meteorological broadcast services, raw data from airport weather stations, raw data from radar stations, and raw data from other official weather stations. All data is processed by OpenWeatherMap in an attempt to provide accurate online weather forecast data and weather maps, such as those for clouds or precipitation. Beyond that, the service is focused on the social aspect by involving weather station owners in connecting to the service and thereby increasing weather data accuracy. The service provides an API with JSON, XML and HTML endpoints and a limited free usage tier. Making more than 60 calls per minute requires a paid subscription. Through the dedicated API, users can request current weather information, extended forecasts and graphical maps and in our case obtain useful weather information that may explain irregular driving behaviour.

\section{Conclusions}
The scope of this paper is to properly present and analyze the components, the development process as well as the overall integration of a driving behaviour monitoring framework prototype designed in compliance with the generic guidelines of Fog computing. Following a brief reference to the fundamental attributes of Fog computing, a high-level architecture description along with the basic components of the prototype were provided. The paper also described the actual implementation and integration process of several sensors, online applications, and third-party modules responsible for cloud data processing and long-term storage. Crucial parts of the final prototype were benchmarked while the overall end-to-end functionality was efficiently presented.

\bibliographystyle{IEEEtran}
\bibliography{references}


\end{document}